\documentstyle[11pt,epsf]{article}
\begin{document}

\setlength{\baselineskip}{15pt}

\begin{flushright}
\begin{small}
\begin{minipage}{8.2cm}
\noindent{\it Proceedings of the 4th Asian
Workshop on First-Principles Electronic Structure
Calculations}
\end{minipage}\vspace*{0.1cm}\\
Taipei, Taiwan (2001)
\end{small}
\end{flushright}

\vspace*{0.2cm}
\begin{center}
\begin{Large}
{\bf
On charge and orbital ordering in La$_{0.5}$Sr$_{1.5}$MnO$_4$\\
}
\vspace{0.5cm}
\end{Large}
Priya Mahadevan$^{1}$, K.Terakura$^{2}$ and D.D.~Sarma$^{3}$ \\
$^{1}$ {\it JRCAT-ATP and AIST, Tsukuba, Ibaraki 305-0046, Japan}\\
$^{2}$ {\it JRCAT-NAIR and RICS-AIST, Tsukuba, Ibaraki 305-0046,
Japan}\\
$^{3}$ {\it Solid State and Structural Chemistry Unit, Indian Institute
of
Science, Bangalore-560012, India}\\
\end{center}

\noindent{\bf Abstract.}\ Using first principle band structure
calculations,
we critically examine results of
resonant x-ray scattering experiments which is believed to
directly probe charge and orbital ordering.
Considering the specific case of La$_{0.5}$Sr$_{1.5}$MnO$_4$, we show
that
this technique actually probes most directly and sensitively small
structural distortions in the system. Such distortions, often difficult
to
detect with more conventional techniques, invariably accompany and
usually
cause
the orbital and charge orderings. In this sense, this technique is
only an indirect probe of such types of ordering. Our results also
provide a
microscopic explanation of the novel types of charge and orbital
ordering
realized in this system and other doped manganites.

In recent times the doped manganites have revealed a wide range of
physical properties resulting from a strong interplay between the spin,
charge
and lattice degrees of freedom \cite{ref1}. In some of the hole-doped
systems, the
charge carriers preferentially occupy certain orbitals, known as
orbital
ordering (OO) or sometimes certain atomic
sites, resulting in charge ordering (CO).
The microscopic origin of these phenomena, especially the charge
ordering
have been explained invoking long-ranged Coulomb interactions
\cite{kontani}.
It is also possible that such orbital and charge orderings are driven
by, or
at least,
substantially assisted by lattice distortions, not considered in a
purely
electronic
mechanism.

It has been suggested \cite{expts,lamno3,lasrmno4}
that resonant xray scattering can directly probe the orbital and charge
orderings. To describe the technique briefly, X-ray diffraction
experiments
are
carried out using photon energies in the vicinity of an absorption edge
to
fufil the resonant condition and the intensities of the diffraction
spots are monitored as a function of the photon energy. While the
intensities
corresponding to the superlattice spots arising from charge or orbital
orderings are
generally very weak, these exhibit remarkable enhancement at the
absorption
threshold, enabling one to detect them easily. This sensitivity in the
diffraction
intensity from the superlattice spots arises from the fact that the
X-ray scattering tensor
is sensitive to various anisotropies such as
orbital ordering and charge ordering.
If a superlattice reflection is identified as arising
due to a certain orbital ordering
occurring at atomic sites A and B, with the energy splitting between
the two
specific
orbitals involved in the ordering equal to $\Delta$, it
has been shown \cite{lamno3}
that the intensity of the superlattice reflection at resonant
conditions
is proportional to $\Delta^{2}$. Similarly at a charge ordering
superlattice reflection, the same dependence on $\Delta$ is seen.
This technique has been extensively
used to probe orbital and charge ordering \cite{expts,lamno3,lasrmno4}
in Pr$_{1-x}$Ca$_x$MnO$_3$,
Fe$_3$O$_4$, LaSr$_2$Mn$_2$O$_7$,
NaV$_2$O$_5$, LaMnO$_3$ and
La$_{0.5}$Sr$_{1.5}$MnO$_4$.

In order to understand the microscopic origin of such orderings, we
have
studied a system, La$_{0.5}$Sr$_{1.5}$MnO$_4$, in which both orbital
as well as charge orderings are present. Mn in this system has an
average
valence of 3.5. X-ray resonant scattering experiments \cite{lasrmno4}
have
been interpreted
to establish that
one of the Mn sites has a valence of 3+, while the other Mn site has a
valence
of 4+, giving rise to an integral valence fluctuation between the two
Mn sites arranged alternately in the lattice. This system appears to be
an
ideal
testing ground, since preliminary crystal
structure data \cite{Immm} did not provide any evidence
for any significant lattice distortion, suggesting a purely electronic
origin for
the charge and orbital orderings. Of course, a purely electronic
mechanism,
{\it via}
electron-electron interactions, will be inaccessible to single-particle
theories; however,
as we shall show, first principle band structure results strongly
suggest
\cite{ourPRL} an alternate
origin of the observed phenomena, while simultaneously establishing
significant
lattice distortions in this system.

We have used the plane-wave pseudopotential method to calculate the
electronic
structure of La$_{0.5}$Sr$_{1.5}$MnO$_4$. Ultrasoft pseudopotentials
were
used for Mn and the La/Sr site. We used virtual crystal
approximation in order to treat alloying effects of La and Sr. The
generalised
gradient approximation \cite{pbe} was used
for the exchange functional, and the calculations
were performed over a k-grid of 4x4x2. The lattice parameters of the
unit
cell
were kept fixed at the experimentally deduced values, and the internal
positions were optimised.

Neutron diffraction experiments \cite{sternlieb} establish that the
Mn spins order below 110~K forming one-dimensional zig-zag
ferromagnetic chains in the ab-plane, coupled antiferromagnetically to
each other (see Fig. 1(a)).
The high temperature structure of the compound is found to be Immm,
with the Mn-O bondlengths equal to 3.66 a.u in the ab-plane, while
the Mn-O bondlength of the apical oxygens was found to be 3.74 a.u. At
217~K,
a charge and orbital ordering transition is found to take place, and
the Mn
sites form a pattern of alternating "Mn$^{3+}$-Mn$^{4+}$" sites in the
ab-plane.
A symmetry lowering was believed to take place, and the experimental
data \cite{sternlieb} were interpreted as arising from a breathing-mode
like
movement of the oxygen atoms in the $ab$ plane towards one set of Mn
atoms,
identified as the higher valent "Mn$^{4+}$" sites. As the structural
information of this compound is shrouded in controversy, we optimised
the
internal positions. This proposed structure is
found to have a higher energy in our calculations. 

The exchange splitting at the Mn site is $\sim$ 3~eV.
Consequently the hopping amplitude between
adjacent zig-zag chains of Mn atoms which are coupled
antiferromagnetically
is small. Hence the electronic structure is governed by these
one-dimensional
chains. Considering a single chain as shown in Fig.~1(a),
it was shown \cite{igor} that the anisotropic
hopping between the Mn $d$ orbitals could explain the OO at
the two Mn sites. We have performed {\it ab-initio} calculations
considering
the
complete 3-dimensional Immm structure \cite{Immm}.
Similar to the findings in ref. \cite{igor}, we also
found the kind of OO that is observed
experimentally. While the $d_{3x^2-r^2}$ orbital was preferentially
occupied
on
Mn(1) atom,
the $d_{3y^2-r^2}$ orbital was preferentially occupied on the other
Mn(3)
atom (see Fig. 1a). These results can be understood
within the framework of a
simple nearest neighbour tight-binding model involving
the $e_g$ orbitals shown in Fig.~1(a): $d_{3x^2-r^2}$ and $d_{y^2-z^2}$ on
Mn(1), $d_{3y^2-r^2}$ and $d_{z^2-x^2}$ on Mn(3) as well as
$d_{3z^2-r^2}$ and $d_{x^2-y^2}$ on Mn(2) and Mn(4) as used in
Ref.~\cite{igor}.
While the $d_{3x^2-r^2}$ orbital at Mn(1) and $d_{3y^2-r^2}$ orbital on
Mn(3) hybridize with both $e_g$ orbitals on Mn(2) and Mn(4), the
$d_{y^2-z^2}$
on Mn(1) and $d_{z^2-x^2}$ on Mn(3) do not. 
\begin{figure}[ht]
\begin{center}
\epsfxsize=400pt \epsfbox{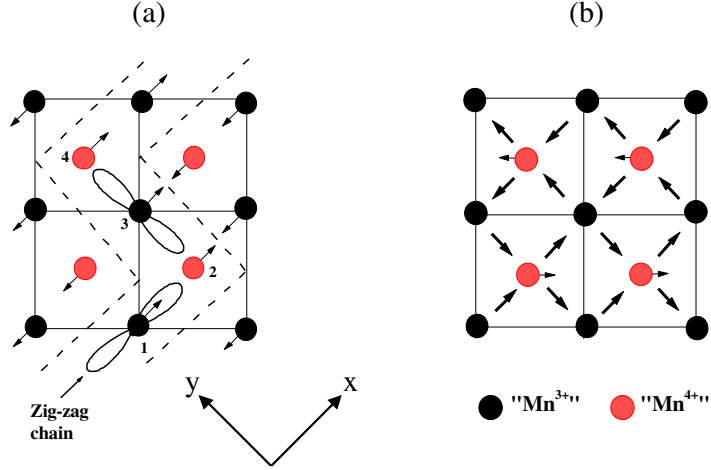}
\end{center}
\vspace*{-1.0cm}
\caption{(a) A schematic figure of the spin, charge and orbital ordering of Mn atoms in the ab plane. The 
magnetic structure consists of zig-zag chains of ferromagnetic Mn atoms coupled antiferromagnetically
to neighboring chain atoms. Dashed lines are drawn parallel to the zigzag chains to highlight each chain.
(b) The direction of displacement of the oxygen atoms are indicated with thick arrows, while small arrows
show the direction of displacement of the Mn atoms. The choice of x and y axes used in the paper is also shown.}
\end{figure}
The eigenvalue spectrum of
such
a simple
model, consists of two bonding bands, energetically separated from four
non-bonding bands, with
the antibonding bands at higher energies. As the average valence of Mn
is
3.5, there are two electrons
in the $e_g$ orbitals of the 4 Mn atoms comprising the chain.
Hence the two bonding bands which have dominantly
$d_{3x^2-r^2}$ character on Mn(1) and $d_{3y^2-r^2}$ character on Mn(3)
are occupied. Thus,
the one-dimensional chains which are a consequence of the magnetic
structure
drive the OO
within these calculations. However, the difficulty in approaching the
problem this way is that
the Immm structure exists only above the
CO/OO temperature and there
are extensive evidence in the literature suggesting a symmetry lowering
below
the ordering transition. However, the exact details of the experimental
crystal structure of this compound is not available, specifically there
is no information on the atom positions.
It was suggested~\cite{sternlieb} that the structure,
within the charge
ordered state,
is Cmmm with lattice constants equal to $\sqrt 2$ a, $\sqrt 2$ a, c
where
a and c are the cell dimensions of the basic Immm cell.
More recent work by Larochelle {\it et al.} \cite{simon}, however,
indicate that the structure
is Ammm or one of its two subgroups Am2m or A222, with
the lattice parameters being $\sqrt 2$a, 2$\sqrt 2$ a,
and c. None of the structural
investigations have been able to provide the atom positions,
particularly
for the oxygen sites, owing to the large unit cell and the existing
data quality \cite{simon}. Ref.~\cite{simon},
however, suggests certain possible
distortions of the oxygens, though not in agreement with
ref.~\cite{sternlieb}.
The structure we find is Bmmm with lattice parameters
$\sqrt{2}$a, 2$\sqrt{2}$a and c. The lattice parameters
that we have obtained are in agreement with those of ref.~\cite{simon,note}.

On optimizing the internal coordinates, the system
exhibited different distortions of the MnO$_6$ octahedra associated
with
different Mn atoms.
The direction of displacement of the oxygen
atoms in the ab-plane is indicated by the arrows shown in Fig.~1(b),
though
the oxygen atoms have been left out of the figure for added clarity.
The Mn site, Mn(3),
showing $d_{3y^2-r^2}$ OO lowered its energy by an elongation
of the Mn-O bonds in the y-direction. This kind of distortion can be
understood easily within the framework of crystal-field effects. A
tetragonal
distortion of a MnO$_6$ octahedron resulting in an elongation of the
Mn-O
bondlengths in the  y-direction, lowers the bare energy of
the $d_{3y^2-r^2}$ orbital. Thus, the sites Mn(1) and Mn(3) behave like
a
"Mn$^{3+}$"
species, sustaining a Jahn-Teller (JT) like
distortion of the surrounding oxygens
in two mutually perpendicular directions, giving rise to the orbital
ordering. As a consequence,
the oxygen atoms surrounding the sites labelled Mn(2) and Mn(4) sustain
distortions
with the oxygen atoms along the x-axis being displaced in one
direction  and the ones along the y-axis being displaced in
another direction, as shown in the figure,
in contrast to the suggestion in ref.~\cite{sternlieb} where
all four oxygen atoms surrounding the Mn(2) and Mn(4) sites
move closer to the Mn atom. The sites Mn(2) and Mn(4)
have been identified as "Mn$^{4+}$" sites in the literature;
within our calculations the charge difference
between the so-called "Mn$^{3+}$" and "Mn$^{4+}$" species is
negligible,
suggesting "charge" ordering to be a misnomer in these cases.

While our calculations suggest
different distortions of the surrounding oxygens about
the "Mn$^{3+}$" and "Mn$^{4+}$"
sites, the actual magnitude of the distortion is fairly
small~\cite{comment1}.
The underestimation of the magnitude of the JT distortion
by these first-principles approaches is well-known, though these
methods do get the nature of distortions
correctly. An example of this
is LaMnO$_3$ where the nature of distortion was
correctly predicted, while
the magnitude of the theoretical JT distortion was
found to be half of the
experimental value~\cite{sawada}.
We simulated the
neutron diffraction pattern using the optimized
coordinates as well as the coordinates assuming a Jahn-Teller
distortion
equal to what is found in LaMnO$_3$. The simulated
diffraction
patterns were nearly identical,
indicating the difficulty of determining the oxygen positions
with any precision from such measurements. It is also
significant that with optimized JT
distortion, the correct ground state magnetic order cannot be
reproduced for LaMnO$_3$~\cite{sawada}.
The situation seems to be the same in the present case,
where the CE type antiferromagnetic state is more stable
than the ferromagnetic state only with the enhanced JT distortion.
Hence in the subsequent analysis, we have extrapolated the magnitude of
the
JT distortion to the value observed in LaMnO$_3$.

The experiments by Murakami
{\it et al.} \cite{lasrmno4} analyzed the energy dependence of the OO
superlattice reflection across the Mn $K$ absorption edge.
The strong energy dependence was interpreted as arising from a
splitting
of $\sim$ 5~eV between the 4$p_x$ and 4$p_y$ PDOS at the sites Mn(1)
and
Mn(3) which
show OO. While the 4$p_x$ states were suggested to be 5~eV
lower than the 4$p_y$ on one Mn atom, the order was reversed on the
other
Mn atom.
\begin{figure}[ht]
\begin{center}
\epsfxsize=250pt \epsfbox{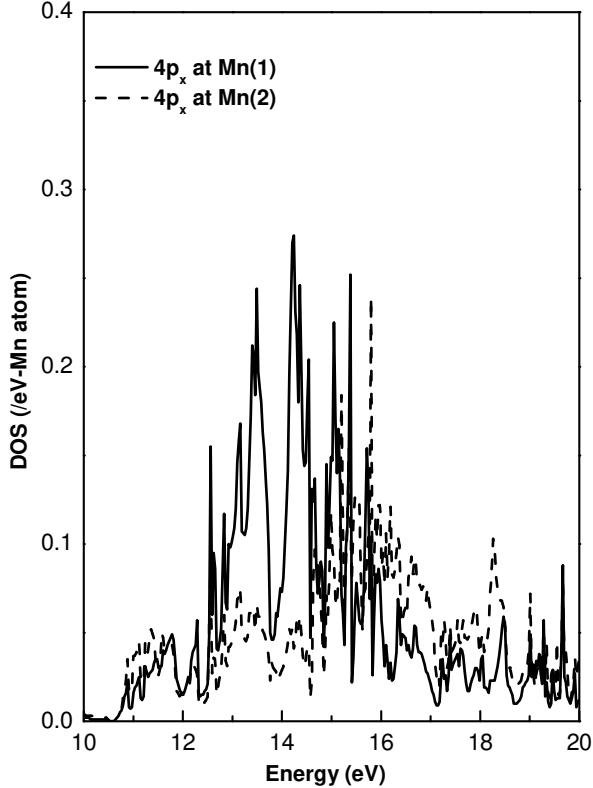}
\end{center}	
\vspace*{-1.0cm}
\caption{ The Mn 4$p_x$ PDOS projected on the sites Mn(1) and Mn(2).}
\end{figure}
Theoretically there are
two contrasting interpretations possible for such a splitting.
Ishihara and Maekawa~\cite{ishihara} argued in the context of LaMnO$_3$
that the splitting in the Mn 4$p$
states induced by the intraatomic $p-d$ Coulomb interaction produces
such a strong tensor character of the scattering form factor.
However, it was pointed out~\cite{other_pap} that
as the 4$p$ states are extended, the suggested $p$-$d$ Coulomb
interaction
strength is unphysically large and the splitting in the 4$p$ states
is actually caused by the JT
distortion, though Coulomb interactions, particularly within the Mn
3$d$
manifold, have important consequences for other properties, such as the
magnetism,
in these systems.
These suggestions are consistent with an earlier analysis of transition
metal
$K$-edge XAS in Fe and Co oxides~\cite{XAS}.
A splitting
of $\sim$ 3~eV is clearly visible in the Mn 4$p$ PDOS
at Mn(1), with the 4$p_x$ states located at lower energies.
The order is reversed at the other Mn site (Mn(3)) which shows
$d_{3y^2-r^2}$
ordering. Thus, our results on La$_{0.5}$Sr$_{1.5}$MnO$_4$ are
consistent
with
the suggestion~\cite{other_pap} in the context of LaMnO$_3$ that the JT
distortion
to be not only the driving force for the orbital ordering, but also
responsible for
the specific experimental effect of the pronounced tensor character of
the
form
factor {\it via} the splitting of the 4$p$ partial DOS.

Strong enhancement in the
intensity at the energy corresponding to the Mn $K$-absorption edge
has been observed at the CO superlattice
reflection.~\cite{lasrmno4}
The experimental results suggest the presence of two types of Mn atoms
in
the system,
with a shift of about 4 eV between the corresponding
absorption edges.
As the absorption edges of formal "Mn$^{4+}$" compounds appear $\sim$4~eV
above the
absorption edge of formal "Mn$^{3+}$" compounds, the authors claimed that
this
was
a direct evidence of an ordering of the two distinct
charge species - "Mn$^{3+}$" and "Mn$^{4+}$" in this system.
From the present calculations, we find that as
the environment for the Mn(1) and Mn(3)
sites is different from the environment for the Mn(2) and Mn(4) sites,
there
is
a substantial modification in the Mn 4$p$ PDOS, explaining the shift in
the
absorption edges, though the net charges
associated with these two sites, 4.53 at Mn(1) and Mn(3), and 4.51 at
Mn(2)
and Mn(4),
are almost identical. As shown in Fig.~2 the 4$p_x$ states
at Mn(1) are found to lie at lower energies compared to
the Mn 4$p_x$ states at Mn(2).
Although there is no substantial charge
difference, the difference in the magnetic moment between the "Mn$^{3+}$"
and the "Mn$^{4+}$" sites is 0.25 $\mu_B$ evaluated within a muffin-tin
radius
of
2~a.u about each Mn atom. A similar, but more pronounced, effect has
been
recently found
in the case of CaFeO$_3$, which was believed to be in a charge
disproportionated state
of Fe$^{3+}$ and Fe$^{5+}$ earlier; it has been
found \cite{cafeo3}
that while the charge state at the
two Fe sites are very similar, there is a strong lattice distortion
that
distinguishes
these two sites and also stabilises considerably different magnetic
moments
at the
two Fe sites.

In conclusion, we have carried out an analysis of the experimental
observations
for orbital and charge ordering in La$_{0.5}$Sr$_{1.5}$MnO$_4$.
Our results indicate the presence of two Mn species with very different
environments. One of the Mn species has a JT distortion of the
oxygen atoms surrounding that atom and hence, in this sense,
may be thought to be similar to
an "Mn$^{3+}$" species. The distortions around the other, so-called
"Mn$^{4+}$", atoms
are different from the simple breathing mode distortion suggested
earlier.
These results establish that the remarkable sensitivity of the
experimental
technique to lattice distortions about the absorbing atom {\it via} the
pronounced shift of
the unoccupied extended states with such distortions, can provide
valuable information which traditional techniques such as neutron
scattering may not be able to provide. This is evident from Fig. 2.
However, it is inappropriate to claim that this technique provides a
{\em direct} probe to the charge and orbital orderings.


\begin{thebibliography}{99}
\bibitem{note} The symmetry found by \cite{simon} is lower than what we find
possibly because to reduce the number of degrees of freedom, we did not allow
the inversion symmetry to be broken during structural optimization.
\bibitem{comment1} In the present structural optimization,
the longer Mn-O distance and the shorter one at sites 1 and 3
are 1.98 {\AA} and 1.90 {\AA}, respectively.  Similarly at sites 2
and 4, they are 1.89 {\AA} and 1.96 {\AA}.
\end{thebibliography}
\end{document}